\documentclass[preprint2]{aastex}

\shorttitle{Lynx}
\shortauthors{Mei et al.}

\begin{document}


\title{Evolution of the Color--Magnitude relation in High--Redshift Clusters: Early--type Galaxies in the Lynx Supercluster at z~$\sim 1.26$}

\author{Simona Mei\altaffilmark{1},
Brad P.~Holden \altaffilmark{2}, 
John P. Blakeslee\altaffilmark{3,1}, 
Piero Rosati\altaffilmark{4},
Marc Postman\altaffilmark{1,5}, 
Myungkook J. Jee\altaffilmark{1}, 
Alessandro Rettura\altaffilmark{4,7},
Marco Sirianni\altaffilmark{5}, 
Ricardo Demarco\altaffilmark{1},
Holland~C.~Ford\altaffilmark{1}, 
Marijn Franx\altaffilmark{6},
Nicole L. Homeier\altaffilmark{1},
Garth D. Illingworth \altaffilmark{2}
}
\altaffiltext{1}{Department of Physics and Astronomy, Johns Hopkins University, Baltimore, MD 21218; smei@pha.jhu.edu, jpb@pha.jhu.edu}
\altaffiltext{2}{Lick Observatory, University of California, Santa Cruz, CA 95064}
\altaffiltext{3}{Department of Physics and Astronomy, Washington State University, Pullman, WA, 99164--2814 USA}
\altaffiltext{4}{European Southern Observatory, Karl-Schwarzschild-Str. 2, D-85748 Garching, Germany}
\altaffiltext{5}{Space Telescope Science Institute, 3700 San Martin Drive, Baltimore, MD 21218}
\altaffiltext{6}{Leiden Observatory, Postbus 9513, 2300 RA Leiden,
Netherlands.}
\altaffiltext{7}{Universit\'e Paris-Sud 11, Rue Georges Clemenceau 15, Orsay, F-91405, France} 

\begin{abstract}

Color--magnitude relations have been derived in two high redshift
clusters at z~$\sim$~1.26.  These clusters, RX~J0849+4452 and  RX~J0848+4453 
(with redshifts of z=1.26 and 1.27, respectively) lie in the highest 
redshift cluster superstructure known today, the Lynx supercluster.
The color-magnitude relation was determined from Advanced Camera for
Surveys (ACS) imaging in the WFC (Wide Field Camera) F775W ($i_{775}$) and
F850LP ($z_{850}$) filters combined with ground--based spectroscopy.
 Early--type cluster candidates have been identified
according to the Postman et al.\  morphological classification. 
In both clusters the bright red early--type population defines a tight color--magnitude relation very similar in color, although the two clusters present different X--ray luminosities and shapes, with RX~J0849+4452 being three times more X--ray luminous and more compact, and having a temperature  two times higher. 
The elliptical galaxy color--magnitude relations (CMR) in RX~J0849+4452 and RX~J0848+4453 show an intrinsic $(i_{775}-z_{850})$ color scatter of $0.026 \pm0.012$~mag and $0.024 \pm 0.015$~mag, respectively, within 2$\arcmin$  ($\sim$~1~Mpc at z=1.26) from the cluster X--ray emission centers. Simple modeling of the scatters using stellar population models from Bruzual and Charlot, gives a mean luminosity--weighted age $\overline t > 2.5$~Gyr ($z_f > $~2.75) and $\overline t > 2.6$~Gyr ($z_f > $~2.8) for ellipticals in RX~J0849+4452 and RX~J0848+4453, respectively. S0 galaxies follow the elliptical CMR; they show larger scatters about the CMR.
 The intrinsic scatter decreases and the CMR slopes are steeper at smaller radii, within both clusters. Within  1$\arcmin$ from the cluster X--ray emission centers, elliptical CMR scatters imply mean luminosity--weighted age $\overline t > 3.2$~Gyr ($z_f > $~3.7).

We conclude that old stellar populations in cluster elliptical galaxies 
are already in place at z=1.26, both in the more evolved cluster
 RX~J0849+4452, and in its less evolved companion RX~J0848+4453. 
Even at a lookback time of 9 Gyr, in the early merging and buildup of
massive clusters, the bulk of the stellar content of the bright elliptical
galaxy population was in place - apparently formed some 2.5~Gyr earlier
at z~$\sim$~3.

\end{abstract}


\def\hst{{\it HST}}

\section{Introduction}

\begin{figure*}
\centerline{\includegraphics[scale=0.7,angle=-90]{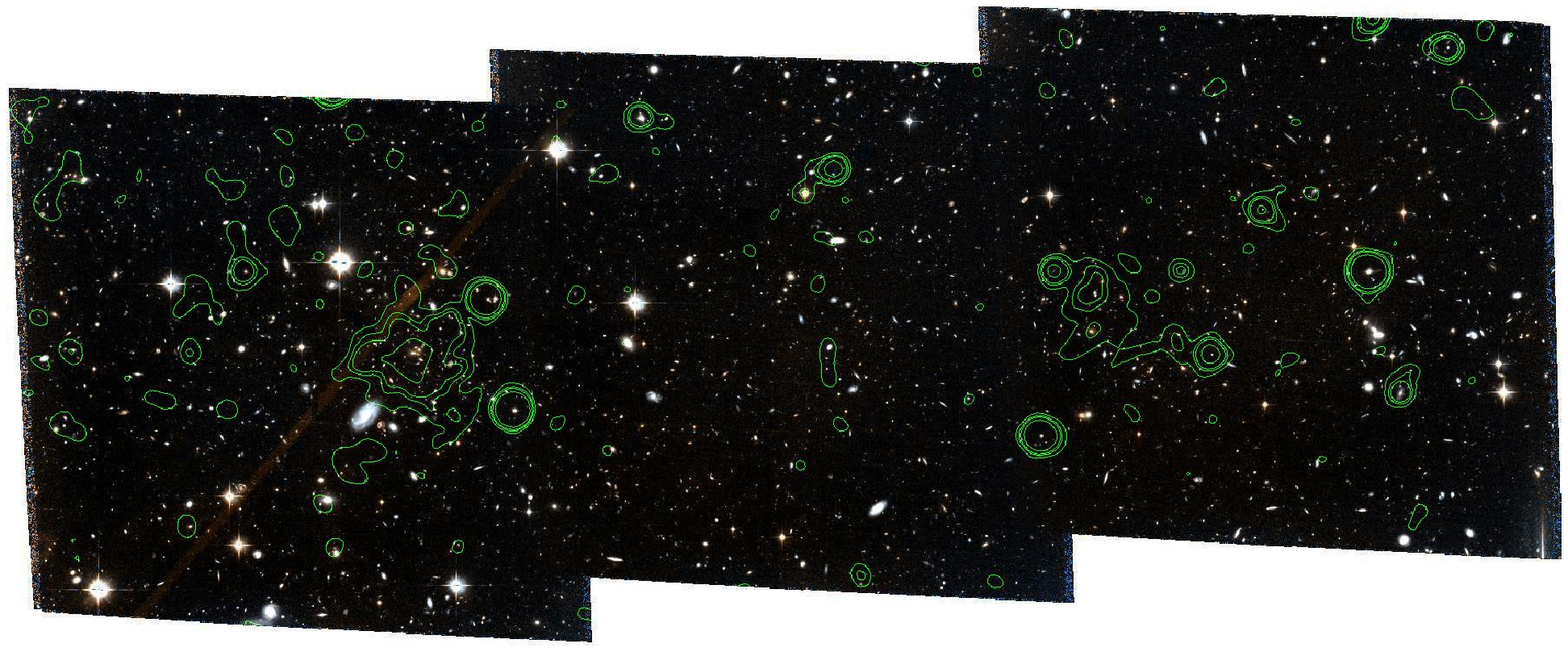}}
\caption {The Chandra X-ray contours overlaid on the ACS color composite image for Lynx E (on the top) and Lynx W (on the bottom). The contours are
adaptively smoothed with a minimum significance of 3 sigma. We refined the alignment of the Chandra image
with respect to the ACS using the X-ray point sources.  {\label{cluster}}}
\end{figure*}

The Lynx supercluster is the highest redshift supercluster
known today (Rosati et al. 1999; Nakata et al. 2005). The two main clusters, RX~J0849+4452 (the Eastern cluster; hereafter Lynx E) and RX~J0848+4453 (the Western cluster; hereafter Lynx W) were detected in the ROSAT Deep Cluster Survey by Rosati et al. (1999), and spectroscopically confirmed respectively at z=1.261 (Rosati et al. 1999) and z=1.273 (Stanford et al. 1997). Lynx W had been previously identified as a galaxy overdensity with $(J-K) > 1.9$ colors down to $K=21$~mag in a near--infrared field survey by Stanford et al. (1997).
The two clusters have a relative projected distance of $\approx$~2~$Mpc$ at z=1.26, in the WMAP cosmology (Spergel et al. (2003): $\Omega_m =0.27$,
 $\Omega_{\Lambda} =0.73$, $h=0.71$), adopted as our standard cosmology. Within a theoretical scenario that predicts hierarchical structure formation, at these redshifts galaxy clusters are still in the process of assembling and the Lynx clusters might be in the process of merging into a more massive structure.
While Lynx E presents a more compact galaxy distribution, with a central bright galaxy merger (van Dokkum et al. 2001, Yamada et al. 2002), eventually leading to a central  cD (cluster dominant) galaxy, the galaxies in Lynx W are more sparsely distributed in a filamentary structure, and do not present an obvious central bright cD galaxy (Fig.~\ref{cluster}). Their X--ray emission from Chandra data (Stanford et al. 2001) confirms the optical distribution, with  Lynx E showing a more compact spherical shape and Lynx W a more elongated one, and luminosities of $L_X^{bol}=(2.8 \pm 0.2) \times 10^{44}$~erg~$s^{-1}$ and $L_X^{bol}=(1.0 \pm 0.7) \times 10^{44}$~erg~$s^{-1}$, respectively (Rosati et al. 1999; Stanford et al. 2001; Ettori et al. 2004). 
Together with a more compact galaxy distribution, this is an indication that Lynx E is likely to be more dynamically evolved with respect to Lynx W. The velocity dispersion for Lynx W was measured to be $\sigma = 650 \pm 170$~km/s (Stanford et al. 2001).
This value is consistent with the recent estimates from Jee et al. (2006) weak lensing analysis of $\sigma = 740^{+113}_{-134}$~km/s and $\sigma = 762^{+113}_{-133}$~km/s for Lynx E and Lynx W, respectively. The most recent estimated of the cluster temperature are $T=3.8^{+1.3}_{-0.7}$~keV and $T=1.7^{+1.3}_{-0.7}$~keV, respectively (Jee et al. 2006 measurements that are consistent with Stanford et al. 2001).
Recently, deep, panoramic multi--color
($VRi',z'$--bands) imaging around these two central clusters identified seven galaxy groups
(Nakata et al.\ 2005) with photometric redshift $z_{phot} \sim 1.26$. This makes the Lynx region 
a unique laboratory, being the only supercluster observed at such a high redshift today, and for this reason, one of the best regions at $z > 1$ in which we can study properties of evolving galaxies within a structure that is still assembling, and in different environments.

The best available optical imaging instrument to study galaxy colors and morphologies at these high redshifts is the Advanced Camera for Surveys (ACS; Ford et al. 2002) on the Hubble Space Telescope (HST), because of its high sensitivity and angular resolution. We have observed Lynx E and  Lynx W as part of our ACS Intermediate Redshift Cluster Survey (guaranteed time observation, or GTO, program 9919). Recent results from this survey show that cluster galaxies at redshift around unity have galaxy distributions, luminosity functions, color and star formation rates similar to local cluster galaxies, although there is already clear evolution in galaxy morphology and ellipticity, with early--type galaxy fractions and axial ratios decreasing with increasing redshift (Blakeslee et al. 2003a;  Demarco et al. 2005; Goto et al. 2005; Holden et al. 2005a; Holden et al. 2005b; Homeier et al. 2005; Postman et al. 2005;Homeier et al., 2006). One of the most universal scaling laws observed in local cluster early--type galaxies is the tight relation between their colors and magnitudes, the color--magnitude relation (CMR; Bower et al. 1992; van Dokkum et al. 1998, Hogg et al. 2004; L\'opez--Cruz et al. 2004; 
Bell et al. 2004, Bernardi et al. 2005; McIntosh et al. 2005). 
This relation was observed up to redshift around unity and shown to evolve back in time in agreement with passively evolving models 
(Ellis et al.\ 1997; Stanford, Eisenhardt, \& Dickinson 1998; van
Dokkum et al. 2000, 2001; Blakeslee et al. 2003a; De Lucia et al. 2004; Holden et al. 2004; Lidman et al. 2004;
Tanaka et al. 2005).
As one of the first results of our ACS cluster survey, Blakeslee et al. (2003a) showed that a tight relation is already in place in the cluster RDS1252.9-292  at $z = 1.24$, from the combination of ACS and ground--based near--infrared imaging and spectroscopic data (see Lidman et al. 2004 for the near--infrared CMR). Because of the high angular resolution and sensitivity of ACS, galaxy colors were measured with a precision unattainable from the ground, and the scatter of the CMR was used to estimate formation ages at $z_f > 2.7$ for the redder cluster elliptical galaxy population (Blakeslee et al. 2003a). 

In this paper, we extend our study to the color--magnitude relation in Lynx E and  Lynx W. The weak lensing mass profile of these two clusters is presented in Jee et al. (2006). The dark mass distribution derived from the weak lensing analysis is in good agreement with the spatial distribution of cluster galaxies and the X--ray emitting gas. 
Our analysis will concentrate on the age of the elliptical population and the galaxy color distribution as a function of distance from the cluster center. We aim to establish whether these two clusters that are so different in X--ray luminosity and baryon distribution, present different early--type populations. If the two populations are similar, it will confirm that early--type galaxies are already in place at redshift larger than unity, both in an evolved and a less evolved cluster, and before these structures might eventually merge in a more extended cluster.

\section{Observations}

\begin{figure*}
\centerline{\includegraphics[scale=0.7]{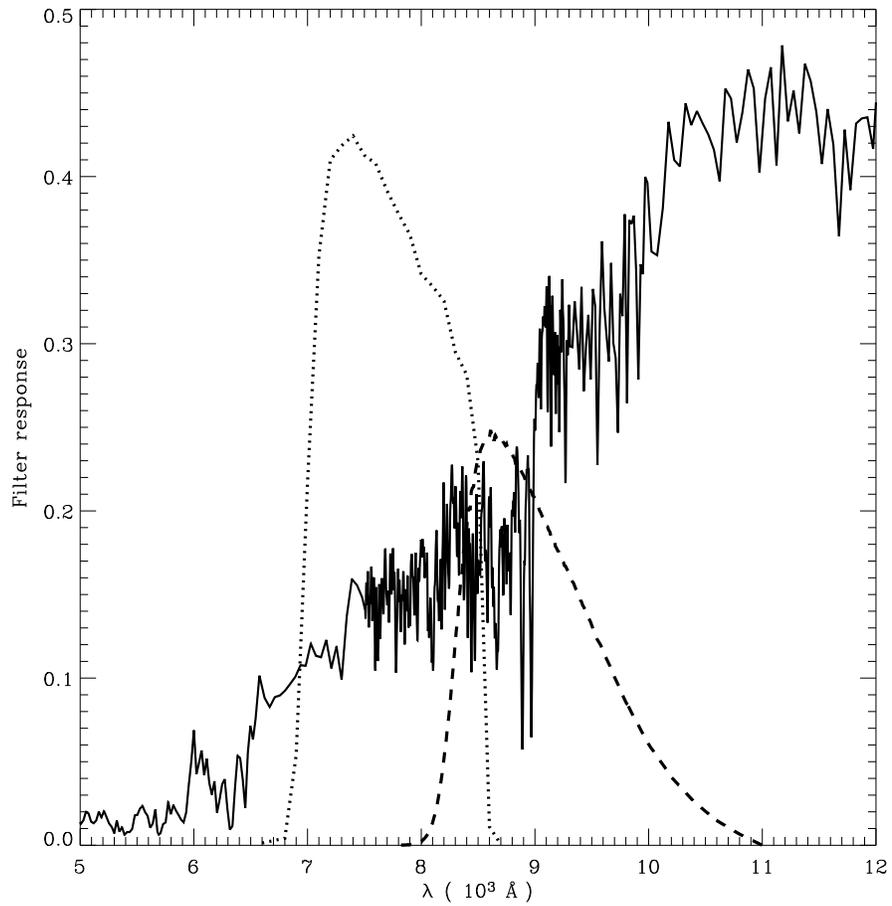}}
\caption {The ACS filter response in the F775W (dotted line) and F850LP (dashed line) bandpasses is shown.
The solid line is a theoretical spectrum from BC03 stellar population models for a galaxy with solar metallicity and age 3.5~Gyr at z=1.27. The spectrum was scaled to arbitrary units to show how the chosen bandpasses sample the light of a model elliptical at this redshift. {\label{filters}}}
\end{figure*}

Lynx E and  Lynx W were
observed in the F775W (from hereafter $i_{775}$) and F850LP (from here after $z_{850}$) bandpasses with 
the ACS Wide Field Camera (WFC) in March and April 2004, for a total of 
22000 and 36500 secs respectively. The two observing filters have been chosen to bracket the 4000~\AA break of a model elliptical galaxy at z=1.26 (Fig.~\ref{filters}).
The ACS WFC resolution is 0.05~\arcsec/pixel, and its  field of view is 210~\arcsec x 204~\arcsec.
 The images were processed with the APSIS pipeline (Blakeslee et al. 2003a, 
2003b), with a {\it Lanczos3}  interpolation kernel.
Our photometry was calibrated to the AB system, with  synthetic photometric zero-points of 25.654~mag and 24.862~mag, respectively in $i_{775}$ and $z_{850}$ (Sirianni et al. 2005). 
A reddening of $E(B-V)=0.027$ was adopted (Schlegel et al. 1998), with $A_{i775}=0.054$ and 
$A_{z850}=0.040$.

To select objects for follow-up spectroscopy, we used the catalog of
Postman et al. (2005) to select potential early-type cluster members.
A combination of ground based imaging in $u$, $g$, $R$, $I_c$ and $z$
was used, in addition to the ACS data, to select additional objects at
the redshift of the cluster.  The data were obtained at the Keck
Telescope with the
 Low Resolution Imaging Spectrograph (LRIS; Oke et al. 1995) and
the Deep Imaging Multiobject Spectrograph (DEIMOS).  The data were
taken over a number of nights using either the 400 or 600 line
mm$^{-1}$ grating with LRIS and the 600 line mm$^{-1}$ grating with
DEIMOS.  The grating tilt was generally set to near 8000 \AA.  With
LRIS, the data were taken with small offsets, $\sim$ 2\arcsec, between
observations to remove the fringing at long wavelengths.  
All of the spectroscopy data were reduced from two--dimensional images to final one--dimensional sky subtracted, wavelength calibrated spectra using a
software package developed by Daniel Kelson.  This uses the technique
outlined in Kelson (2003) for sky subtraction.  For this paper, the
redshifts were measured from the centroid of absorption or/and
emission lines.  Redshifts are available for 20 galaxies out of the above selected candidates, from published data (9 from Stanford et al. 1997, 8 from Rosati et al. 1999) and the two spectroscopic runs at the Keck telescope described above.

\section{Object selection and photometry}

Following Blakeslee et al. (2003b), for a first choice of cluster candidates our first photometry was performed using SExtractor (Bertin \& Arnouts 1996) in
{\it dual-image mode}, as in Ben\'{\i}tez et al. (2004). This means that after object fluxes were first measured independently in the two filters, object detection employed the two filters simultaneously (an object is assumed as detected only if it is detected in both bands). We used the resulting catalog of detections and colors, combined with the morphological classification
from Postman et al. (2005),  to select potential
early-type cluster members. 
 Our first sample included early--type galaxies (ellipticals, S0 and S0/a) with $0.5 <(i_{775} - z_{850}) < 1.2$, inside a radius of $2\arcmin$ from each cluster center, taken as the center of the cluster X--ray emission from Stanford et al. (2001). 
 A scale of two arcminutes correspond to 
$\approx 1 Mpc$ at $z=1.26$ in the WMAP cosmology.  
From this sample, 43 (23 ellipticals, 14 S0s, and 6 S0/as) and 30 (10 ellipticals, 5 S0s, and 15 S0/as) objects, respectively for Lynx E and  Lynx W, were identified.
To avoid systematics because of internal galaxy gradients, our final colors
were measured within galaxy effective radii ($R_e$), following the approach in
Blakeslee et al. (2003b) and van Dokkum et al. (1998, 2000). $R_e$ is the  half
light radius along the major axis of the galaxy best fitting model.  
The major effect of internal galaxy gradients on this sample would be a steepening of the CMR slope (e.g. by $\sim$50\% when isophotal colors from Sextractor are used).
 $R_e$ values were
derived with the program GALFIT (Peng et al. 2002), constraining the {\it Sersic} index to $n \leq $ 4.
Since the PSF is $\approx 10\%$ broader in the $z_{850}$ band, each galaxy image was
deconvolved using the CLEAN algorithm (H{\"o}gbom et al. 1974) in order to remove
blurring effects.
$(i_{775} - z_{850})$ colors were measured on the deconvolved images within a 
circular aperture equal to $R_e$. When $R_e < 3$ pixels, we have set it equal to 3 pixels. Our median $R_e$ is $\approx$~5~pixels.
Errors in the colors were estimated by adding in quadrature to the uncertainties in flux, the uncertainty due to flat fielding, PSF variations, and pixel-to-pixel correlation for ACS (Sirianni et al. 2005).
This last total uncertainty was estimated by measuring the standard deviation of
photometry in the background for circular apertures in the range of the measured effective radii.
Errors in colors are between 0.02 and 0.08~mag, and are dominated by the Poissonian noise in the galaxy fluxes.
Among the first selected 73 galaxies, we selected 46 (27 in Lynx E and 19 in Lynx W) cluster member candidates 
with colors within $R_e$: $0.8 <(i_{775} - z_{850}) < 1.1$, $z_{850}$ fainter than 21~mag and distances within two arcminutes from the two cluster centers. 14 of them ($\approx$~70\%) are confirmed cluster members, 6 are interlopers (two with $z \approx$~1.06, one with $z \approx$~0.9 and three with $z \approx$~1.14).  We are left with 14 confirmed members and 26 cluster member candidates, for a total of 40 galaxies.

\section{Results}

In this section we will discuss the color--magnitude relation for each cluster separately and then compare them. 


\subsection{Fit to the Color-Magnitude relations}

We fitted the following linear color--magnitude relation to various galaxy
subsamples:
\begin{equation}
i_{775} - z_{850} =  c_0 + Slope  (z_{850} - 22.5)
\end{equation}
where $c_0$ is the CMR zero point and $Slope$ the CMR slope.

\begin{figure*}
\centerline{\includegraphics[angle=90,scale=0.7]{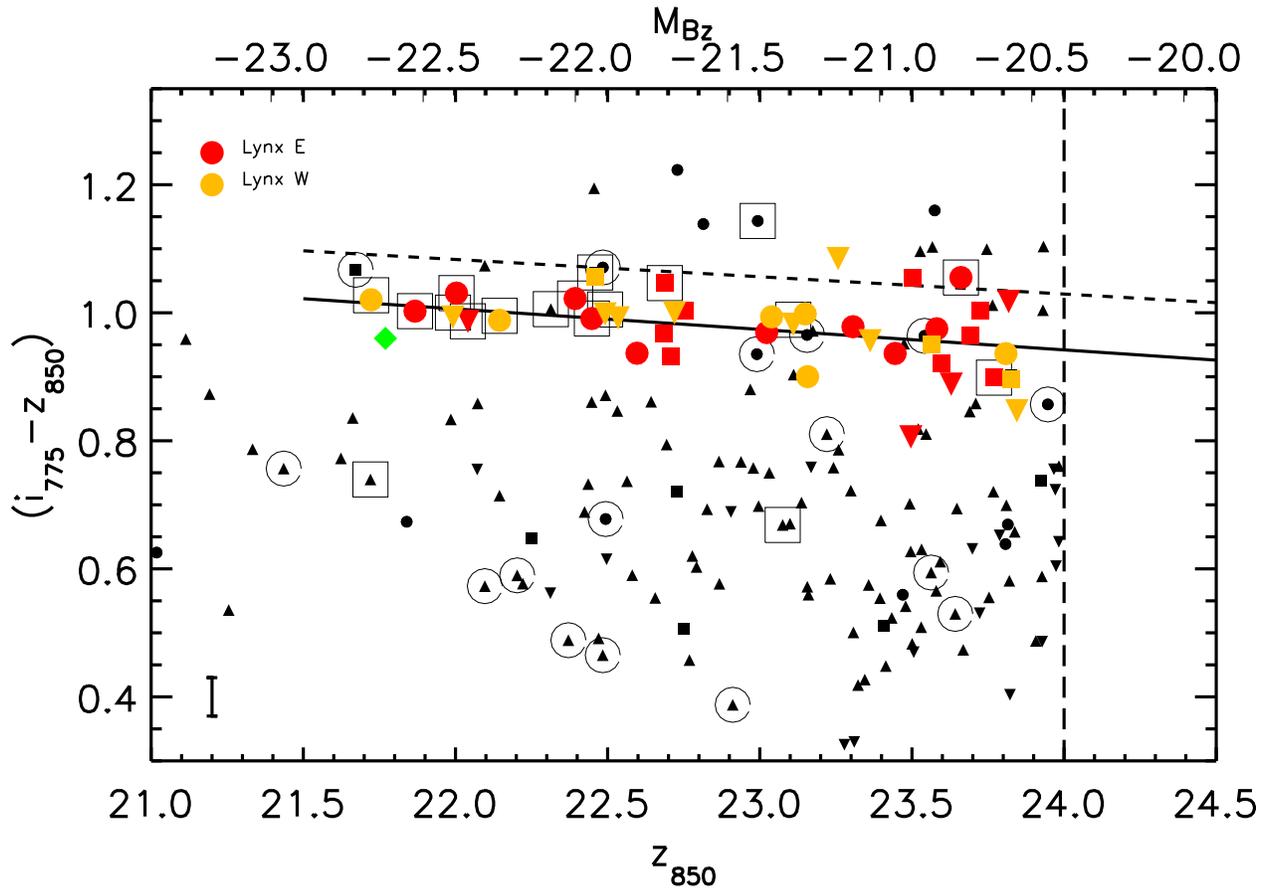}}
\caption {The color-magnitude relations for Lynx E and  Lynx W. 
Elliptical galaxies are shown as circles, S0 as squares, S0/a as upside down triangles, spirals as triangles. Red sequence selected objects are red and orange for  Lynx E and Lynx W, respectively. Open circles are plotted around confirmed interlopers. Boxes are plotted around spectropically confirmed members.  Colors in  Lynx W have been shifted by +0.007~mag to account for the small difference in redshifts, using Bruzual \& Charlot (2003) stellar population model predictions. The green diamond is the three galaxy merger in Lynx W.  Lynx E and Lynx W show very similar CMR scatters and zero--points. The continuous line shows the fit to the elliptical CMR from both clusters.  The dashed line is the color--magnitude relation for the Coma cluster K-corrected to z=1.26, and scaled to these bandpasses. The rest frame $M_B$ is shown on the upper axis. The median color error is shown in the bottom left.
{\label{cmd}}}
\end{figure*}

Fig~\ref{cmd} shows the color-magnitude relations for Lynx E and  Lynx W. 
Circles and squares are used for elliptical and S0 galaxies, respectively. Upside down triangles are S0/a. Triangles are late type galaxies. Boxes are plotted around confirmed cluster members. Open circles are plotted around confirmed interlopers. Red and orange are used for CMR early--type galaxies in Lynx E, and Lynx W, respectively. 
Colors in  Lynx W have been shifted by 0.007~mag to take into account the small difference in redshift, using Bruzual \& Charlot (2003; BC03) single burst solar metallicity stellar population models. 
The color-magnitude relation was fitted using a robust linear fit based on Bisquare weights (Tukey's biweight; Press et al. (1992)), and the uncertainties on the fit coefficient were obtained by bootstrapping on 10,000 simulations. The scatter around the fit was estimated from a biweight scale estimator (Beers, Flynn \& Gebhardt 1990), that is insensitive to outliers, in the same set of bootstrap simulations.
A linear least square fit with a three-sigma clipping, and standard rms scatter give similar results within $\approx 0.001-0.002$~mag in slope and in scatter. 
To estimate the internal galaxy scatter, 
we adopted two strategies: first, we subtracted in quadrature from this last value the scatter due to the average galaxy color error, and second we calculated the internal scatter for which the $\chi^2$ of the fit would be unity.
The two estimates of the internal scatter agree
within 0.001~mag.

Independent fits to the CMR have been obtained for ellipticals, S0s and all early--types together, in the two clusters. S0/a galaxies were also added to the analysis, since most of them are observed to be in the process of merging to possibly form red sequence galaxies (see below the colors and magnitudes of a merger of three S0/a galaxies in Lynx W, that lies on the red sequence). Our results are shown in Table~\ref{results} at
1$\arcmin$ ($\approx$0.5~Mpc), 1.5$\arcmin$  ($\approx$0.8~Mpc), and 2$\arcmin$  ($\approx$1~Mpc) distance from the cluster center.

\subsection{Color-Magnitude relation in Lynx E}

The fit to the color--magnitude relation for E+S0 
in Lynx E within 2$\arcmin$ (19 early--type galaxies) is:
\begin{equation}
(i_{775} - z_{850}) = (0.99 \pm 0.01) + ( -0.020 \pm 0.018) \times (z_{850} - 22.5)
\end{equation}
with an internal scatter of $0.038 \pm 0.008$~mag. 
Selecting galaxies within 1$\arcmin$ from the cluster center, the slope  of the E+S0 sample gets steeper to a value of $-0.040 \pm 0.015$ and the scatter decreases to $0.020 \pm 0.006$~mag.

When only ellipticals are selected, the slope and the scatter are, respectively, $-0.024 \pm 0.020$ and $0.026 \pm0.012$~mag within 2$\arcmin$. 
From the scatter around the CMR, and with simple modeling using stellar population models, galaxy ages can be estimated, following the approach from van Dokkum et al. (1998) and Blakeslee et
al. (2003a) (see also Blakeslee et al. 2006 and Mei et al. 2006), with a precision unattainable with ground--based data. We consider two simple models: the first model is a  {\it single burst} model,
in which galaxies form in a single burst at time $t_f$, randomly chosen to be between a time $t_{end}$ and a time $t_0$, corresponding to the recombination epoch.
The second model is a
model with continuous, constant star formation rate ({\it constant star formation}) over a range of time between $t_1$
and $t_2$,  both randomly chosen to be between a time $t_{end}$ and
$t_0$. For each $t_{end}$, colors were simulated 
for 10,000 galaxies with formation 
ages (single burst or constant) varying between $t_{end}$ and 
$t_0$. From those simulations we derived color scatters as a function of
$t_{end}$. For each $t_{end}$ we estimated a mean luminosity--weighted age, obtained by luminosity weighting galaxy ages between $t_{end}$ and $t_0$.

Assuming a simple BC03 {\it single burst} solar metallicity model, and comparing the simulated scatter as a function of  $t_{end}$ to the scatter measured around the fit of the CMR, we derive ages $>0.8$~Gyr ($z >1.6$), with mean luminosity--weighted age $\overline t = 2.6$~Gyr ($z_f \approx$~2.8) for the elliptical galaxies within 2$\arcmin$, and ages  $>2.6$~Gyr ($z >2.8$), with mean luminosity--weighted age $\overline t = 3.4$~Gyr ($z_f \approx$~4.2) within 1$\arcmin$.
When a {\it constant star formation} solar metallicity model is considered, we obtain ages $>0.3$~Gyr ($z >1.4$), with a mean luminosity--weighted age $\overline t = 2.5$~Gyr ($z_f \approx$~2.8)  for the elliptical galaxies within 2$\arcmin$, and ages $>2.2$~Gyr ($z >2.5$), with a mean luminosity--weighted age $\overline t = 3.4$~Gyr ($z_f \approx$~4.2) within 1$\arcmin$.

From the two very simple models above, we obtain a mean luminosity--weighted age $\overline t > 2.5$~Gyr ($z_f > $~2.8); that is at $t$ less than 2.5~Gyr after the origin of the Universe) for ellipticals in Lynx E,  within 2$\arcmin$ from the cluster center. 
If we consider the central region of the cluster,  within 1$\arcmin$, we obtain a mean luminosity--weighted age $\overline t > 3.4$~Gyr ($z_f > $~4.2); that is at $t$ less than 1.5~Gyr after the origin of the Universe).

When S0 and S0/a are added to the sample, zero points and slopes do not change, but the scatter about the CMR increases, implying mean ages younger of $\sim$~0.3~Gyr (adding S0) and $\sim$~0.5~Gyr (adding S0 and S0/a).

\subsection{Color-Magnitude relation in Lynx W}

Nine elliptical and S0 galaxies were selected in Lynx W. The fit to the E+S0 sample shows
a steeper slope in the CMR than for Lynx E, but with a larger statistical error:
\begin{equation}
(i_{775} - z_{850}) =(1.00 \pm 0.01) + ( -0.056 \pm 0.018) \times (z_{850} - 22.5)
\end{equation}
and an internal scatter of $0.027 \pm 0.015$~mag.

The elliptical galaxies show a shallower slope, ($-0.043 \pm 0.031$), but still larger than the slope of the elliptical CMR in  Lynx E.
The internal scatter of the elliptical CMR is $0.024 \pm 0.023$~mag, within $2 \arcmin$ from the cluster center, and decreases to $0.017 \pm 0.028$~mag within $1 \arcmin$.
These scatters give ages $>0.95$~Gyr ($z >1.6$), with mean luminosity--weighted age of $\overline t = 2.6$~Gyr ($z_f \approx$~2.9) within  $2 \arcmin$,  and 
ages $>2.3$~Gyr ($z >2.5$), with mean luminosity--weighted age of$\overline t > 3.3 $~Gyr ($z_f \approx$~3.9) within $1 \arcmin$,for a simple BC03 {\it single burst} solar metallicity model.
Using a {\it constant star formation} solar metallicity model, for elliptical galaxies within 2$\arcmin$ and within 1$\arcmin$, ages  $>0.4$~Gyr ($z >1.4$), with mean luminosity--weighted ages of $\overline t = 2.6$~Gyr ($z_f \approx$~2.8), and ages  $>0.1.6$~Gyr ($z >2$), with mean luminosity--weighted ages of $\overline t = 3.2 $~Gyr ($z_f \approx$~3.7) are obtained, respectively.

From above, the mean luminosity--weighted age obtained for elliptical galaxies in Lynx W is $\overline t > 2.6$~Gyr ($z_f \approx$~2.8), within 2$\arcmin$ from the cluster center. We obtain mean luminosity--weighted ages  $\overline t > 3.2$~Gyr ($z_f \approx$~3.7), within 1$\arcmin$ from the cluster center.

As in Lynx E, adding to the sample S0 and S0/a does not change zero points and slopes, but increases the scatter about the CMR, with estimated mean ages younger of $\sim$~0.2~Gyr (adding S0) and $\sim$~0.3~Gyr (adding S0 and S0/a). 

Van Dokkum et al.'s (2001) color--magnitude relation for early--type galaxies in this cluster show a slope ($-0.02 \pm 0.03$) in the $U-V$ rest--frame, 
 using ground based BRIzJK imaging combined with WFPC2 and NIC3 morphologies. In the $(i_{775} - z_{850})$ color this slope corresponds to $-0.066$~mag (using BC03 single burst solar metallicity, age 4~Gyr stellar population models), consistent both with the overall (E+S0+S0/a) slope of $-0.051 \pm 0.014$ and the E+S0 slope of $-0.056 \pm 0.014$  that we found.

\subsection{Color-Magnitude relation in the combined cluster sample}

The continuous line in Fig~\ref{cmd} is the fit to the color--magnitude relation for E galaxies within $2\arcmin$ in the combined cluster sample:
\begin{equation}
(i_{775} - z_{850}) = (0.99 \pm 0.01)+ ( -0.031 \pm 0.012) \times (z_{850} - 22.5)
\end{equation}
which, within the errors, is equal to the fit for ellipticals in each cluster.
The internal galaxy scatter around this fit is 0.025~$\pm$0.010~mag.
When considering different regions from the cluster center, between 1$\arcmin$ to 2$\arcmin$, the scatter around the CMRs increases (always within 1~$\sigma$ though) adding the external regions of the clusters, while zero points are similar, suggesting that galaxies closer to the cluster center are an older elliptical galaxy population or that there are more interlopers within 2~$\arcmin$.  Contamination by foreground galaxies was shown to be high ($\approx$~30\%) in the spectroscopic sample that we used (Stanford et al. 1997; Rosati et al. 1999; Holden et al. in preparation).

\begin{figure*}
\centerline{\includegraphics[scale=0.7]{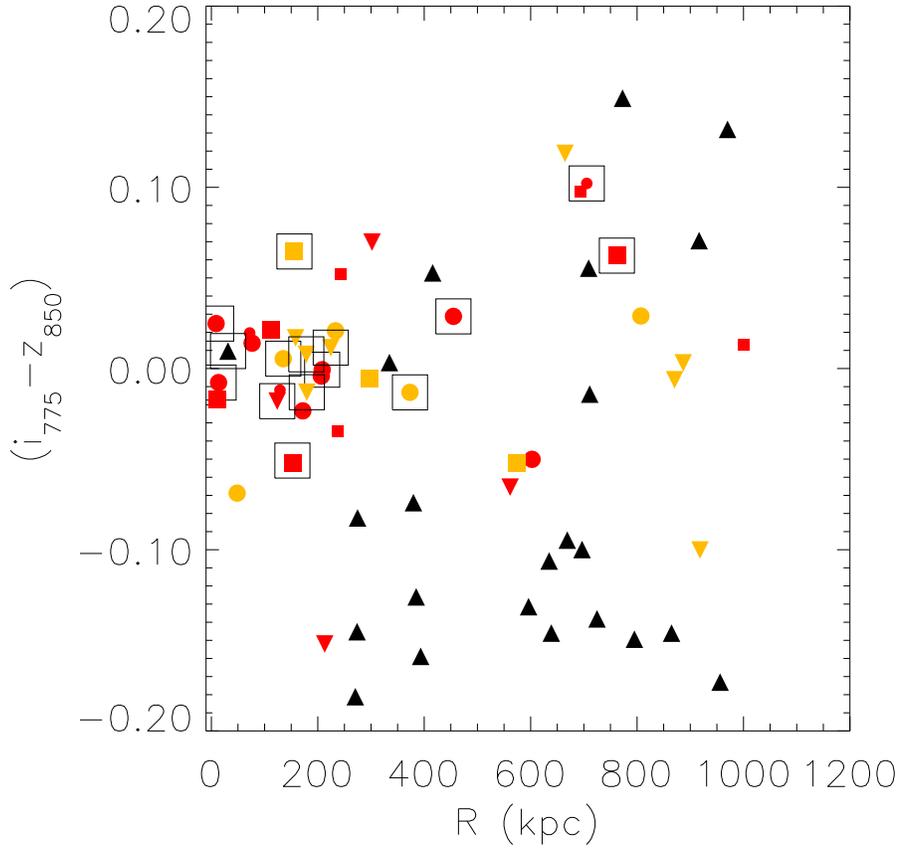}}
\caption {$(i_{775} - z_{850})$ colors (corrected by the elliptical CMR from both clusters) as a function of distance from the cluster centers $R$ for CMR galaxies ($(i_{775} - z_{850})$ between 0.8 and 1.1~mag). As in Fig.~\ref{cmd}, elliptical galaxies are shown as circles, S0 as squares, S0/a as upside down triangles, spirals as triangles. Red and orange colors for  Lynx E and Lynx W, respectively. Interlopers are not shown.  Smaller symbols show galaxies with $z_{850}$ fainter than 23.5~ mag. Similar distributions of colors as a function of distance from the cluster center are observed for the two clusters. {\label{colkpc}}}
\end{figure*}

Fig~\ref{colkpc} shows galaxy colors as a function of distance from the cluster centers, once they are corrected by the CMR fit to the combined sample. Most (70\%) of the CMR early--type galaxies lie close to the cluster center, within 1$\arcmin$ ($\approx$~500~kpc). Of those, 50\% are ellipticals. 
Similar distributions of colors as a function of distance from the cluster center are observed for the two clusters. Once the colors are corrected by the CMR, the average color does not depend on the distance from the cluster center, while the scatter around the mean increases slightly with distance.

\subsubsection{CMR Slope and Scatter}

The slope of the early--type CMR from the fit above is consistent with previous results for the Coma cluster and for RDCS~1252--2927 from Blakeslee et al. (2003b) that we show in Fig~\ref{cmd}. The dashed line is the color--magnitude relation for the Coma cluster scaled to our colors and redshift, without evolution. The slope of the CMR in the Coma cluster, transformed to our band--passes at $z=1.27$, is $-0.027$.
The dotted line is the CMR derived by Blakeslee et al. (2003), for ellipticals in RDCS~1252--2927, and the dashed--dotted line is the CMR for ellipticals in RXJ~J0910+5422, both scaled to z=1.26 without evolution. 

The S0 CMR has steeper slope and similar zero point with respect to the ellipticals. The S0 CMR slope becomes much steeper in the 1~$\arcmin$ cluster regions ($-0.063 \pm 0.035$), while the elliptical galaxy slope is slightly steeper ($-0.036 \pm 0.011$). The scatter around the S0 CMR fit is larger than the ellipticals, independent of the distance from the cluster center.

When we compare the Lynx E and Lynx W CMRs, we observe similar scatters, and consequently, when assuming a simple population model, similar ages for elliptical galaxies in the two clusters. However, Lynx W CMR slope is steeper than in Lynx E, for all early--type populations, even if it is statistically different only for the E+S0 samples for distances from the cluster centers larger than 1.5~$\arcmin$.

Different slopes in a CMR are due to a different evolution of galaxies of different masses. The CMR is thought to be due to a mass--metallicity relation (Kodama \& Arimoto 1997; Gladders et al. 1998; Kodama et al. 1998), and CMR slopes change with time, since more metal--rich galaxies become redder faster than metal--poor galaxies. Steeper slopes would indicate older galaxy ages.
This is true if the brightest and the faintest galaxies evolve at the same time. However, in recent studies of galaxies in clusters, the brightest galaxies seem to be already evolved, and have ended their star formation, at z~$\ge$~1, while fainter galaxies are still evolving and forming stars ({\it down--sizing}; Cowie et al. 1996).
Our results are consistent with this scenario.
The steeper slopes for Lynx W seem due to a lack of red, faint early--type galaxies in this cluster, which is observed in Lynx E (for  distances from the cluster centers larger than 1.5~$\arcmin$). The  Lynx W early--type $(i_{775} - z_{850})$ colors lie well on the Lynx E red sequence for magnitudes brighter than $z_{850}=23.5$~mag (corresponding to an absolute rest--frame B magnitude of $M_B = -20.9$~mag). However, at magnitudes  $z_{850}$ fainter than $23.5$~mag, all four early type in Lynx W lie below the Lynx E red sequence, and two of the them (the S0 and the S0/a) lie at more than 500~kpc from the cluster center. 
The lack of a faint red population is also observed in other works. In Lynx W, we observe it just at the limiting magnitude of our morphological classification and it might be real or just due to a statistical fluctuation.
Tanaka et al. (2005) studied the color--magnitude relation in a local sample of clusters from the SDSS (Sloan Digital Sky Survey; York et al. 2000), and the two clusters CL0016+1609 (z=0.55) and  RXJ0152.7--1257 (z=0.83). These results support a scenario in which giant galaxies (extrapolating their model at z=1.26 those would be galaxies brighter than $M_B = -20.2$~mag) complete their star formation before z~$\sim$~1, while small objects continue to evolve. In fact, the brightest galaxies in their highest redshift cluster, RXJ0152.7--1257, are already in place, while the faint--end of the galaxy population is still in the process of building--up, e.g. their scatter around the main sequence is larger than for the brightest population, and their fraction is smaller. This last phenomenon, the lack of faint galaxies in the red sequence is also called {\it truncation of the red sequence}.
A truncation of the red sequence was already pointed out by Kajisawa et al. (2000) and Nakata et al. (2001), in the 3C~324 cluster at z$\sim$1.2.
De Lucia et al. (2004) observed a deficit of faint (absolute rest--frame B--magnitude $M_B$ fainter than -19.9~mag) red galaxies in their sample of 4 clusters at $z\sim 0.8$ with respect to Coma and 2dF (The Two Degree Field) galaxy survey clusters. They suggested that it implies that a large fraction of the red faint population in local clusters moved to the red sequence only in recent times.  

\begin{figure*}
\centerline{\includegraphics[scale=0.7]{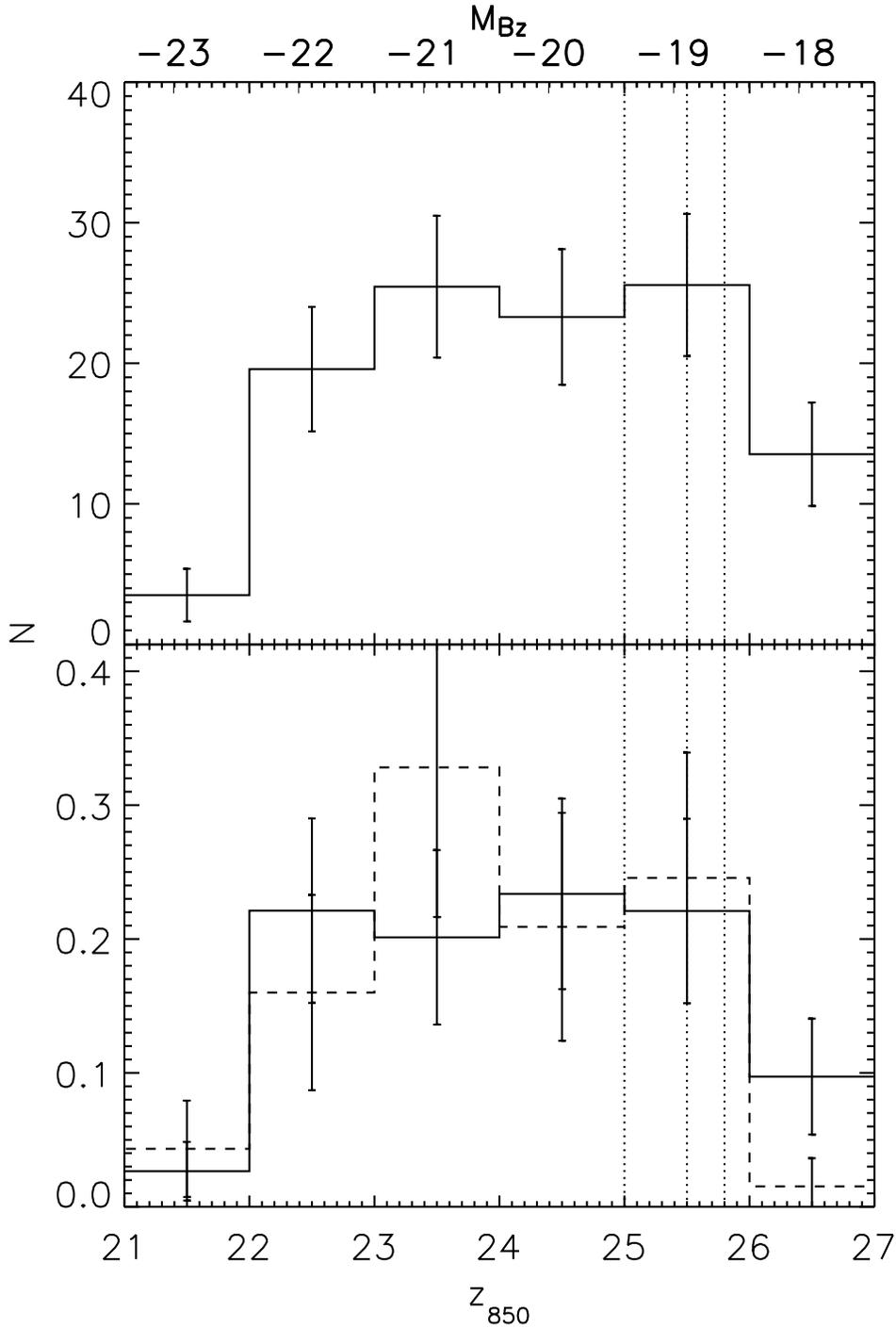}}
\caption {On the top panel, is shown the histogram of the galaxies within 3--$\sigma$ from the total elliptical CMR. Errors are calculated as Poissonian. The dotted lines show the magnitudes corresponding to detection completeness of 80\%, 50\% and 30\% (from left to right; see text for details). On the bottom, the same for galaxies belonging to Lynx E (continuous line) and Lynx W (dashed line). Rest frame $M_B$ is shown on the upper axis.  {\label{trunc}}}
\end{figure*}

Our Lynx sample is not deep enough to estimate if there is a truncation in the red sequence. 
In Fig.~\ref{trunc} is shown the histogram of all galaxies within 3--$\sigma$ from the total elliptical CMR. This sample has been selected from the original Sextractor sample described in Section 3, before selecting early--type galaxies.
We obtain a total of 217 galaxies within 2~$\arcmin$ from the two cluster centers.
Photometric redshifts were obtained with the BPZ (Bayesian Photometric Redshifts) software (Benitez 2000), combining our ACS photometry with V and $R_c$--band photometry from public Subaru/Suprime-cam images of this field. Galaxies with photometric redshift between 0.8 and 1.5 (consistent with an error of $\sim$~0.3~mag in photometric redshifts around z=1.2) were selected (169 galaxies; 80\% of the galaxies within 3--$\sigma$ from the total elliptical CMR).
The histogram obtained for these galaxies was then corrected by the expected number counts in the field, using as a control region the GOODS-S 
(Great Observatories
Origins Deep Survey--South; Giavalisco et al. 2004) ACS field
observed in the same filter and with similar magnitude limits as our cluster field. The dashed lines in Fig.~\ref{trunc} show our magnitude limits at 80\%, 50\% and 30\% completeness.
The completeness has been estimated by simulating 5,000 galaxies on our ACS image and detecting them with the same technique described in Section~3. In the detection we required a match within 3 pixels in the spatial coordinates, and within 0.5~mag in magnitude. We have used three ellipticals chosen among our CMR galaxies as templates for the simulations. The sample has a limiting magnitude of $z_{850}=25$~mag (equivalent to absolute rest frame magnitudes $M_{B}=-19.4$~mag and $M_{V}=-20.10$~mag from BC03 models), corresponding to a completeness in detection of 80\%. Our sample probes the dwarf galaxies regime ($M_B$ fainter than -20.2~mag from Tanaka et al. 2005) too much close to our magnitude limit of $z_{850}=25$~mag to be able to draw firm conclusions about a truncation of the red sequence. However, we can observe a tendency of the number counts to be at least constant for magnitudes brighter than $z_{850}=26$~mag ($M_{B}=-18.4$~mag; $M_{V}=-19.10$~mag). Deeper observations are needed to further address this issue.

We can conclude that the bright red early--type galaxy population is already in place in both clusters at z~$\sim$~1.26.

\subsubsection{Color-Magnitude relation zero points}

Lynx E and Lynx W CMR have very similar zero points. From the combined early--type CMR,  $(i_{775} - z_{850}) = 0.99 \pm 0.03$~mag between $z_{850}= 21.5$~mag to 24~mag.
 From the simple model simulations described above for both the {\it single burst} and the {\it constant star formation} solar metallicity model, the mean galaxy color corresponding to our derived mean luminosity-weighted ages is $(i_{775} - z_{850}) = 0.92 \pm 0.03$~mag.  The observed $(i_{775} - z_{850})$ colors are redder (by $0.07 \pm 0.04$~mag) than expected from simple BC03 stellar population models.

Since we are already considering old galaxy populations, the Lynx  cluster galaxies should have on average twice solar metallicities,  
or should all be reddened by dust, in order to explain these redder colors in terms of passive evolution models




However, an offset in the $(i_{775} - z_{850})$ color might also be explained by an uncertainty of a few hundreds of a magnitude arising from the uncertainty in the shape of the ACS $z_{850}$ bandpass response (see discussion in Sirianni et al. 2005).
The ACS bandpass responses have been calibrated from the observed  versus the predicted ACS count rates, up to 9000\AA. This calibration permits us to estimate ACS bandpass zero points with an uncertainty of 0.01~mag.
The local 4000\AA~break observed in  galaxy templates with age 4~Gyr and solar metallicity is redshifted around 9000\AA~at z=1.26. This means that most of the galaxy light at z=1.26 lies at wavelengths larger than 9000\AA, where the ACS $z_{850}$ bandpass calibration is more uncertain.
We have simulated variations in the ACS $z_{850}$ bandpass response function, adding to the  $z_{850}$ response function an exponential plus linear tail for wavelengths larger than 9000\AA. While keeping the AB zero point derived with the new response functions within 0.01~mag of the zero point in Sirianni et al. (2005), we obtain a maximum offset of $\approx$~0.05~mag in the zero--point of an early--type template spectrum from BC03 (Age 4~Gyr and solar metallicity). 
In a second test we used the spectroscopically selected 
catalog of galaxies  (Vanzella et al. (2005)) in the Chandra Deep Field South from the GOODS-S (Great Observatories
Origins Deep Survey--South; Giavalisco et al. 2004) sample, observed from rest-frame UV to rest-frame near-IR. We selected galaxies with redshift determination in the range $1.2 < z < 1.3$, obtaining a sample of 20 objects. To obtain SEDs of these objects, we used broad-band photometry in the observed ACS $B_{435}$, $V_{606}$, $i_{775}$, $z_{850}$ bands (Giavalisco et al. 2004),  
GOODS/VLT/ISAAC images in the $J$ and $Ks$ filters (Vandame et al., in 
preparation),
and first epoch GOODS/Spitzer/IRAC images in the
 $ch1_{3.5 \mu m}$, $ch2_{4.5 \mu m}$, $ch3_{5.8 \mu m}$, $ch4_{8.0 \mu m}$ observing bands (Dickinson et al., in preparation), aperture--matched. The single SEDs (Spectral Energy Distributions) were fitted by stellar population models, with and without the $z_{850}$ measurement (Rettura et al., in preparation). Comparing the measured $z_{850}$ with the prediction from the SED fit, we estimate a maximum average offset of $ 0.05 \pm 0.15$~mag in the ACS $z_{850}$ zero point. Either test can only allow us to estimate an approximate upper limit to a possible offset in the zero point.
A refinement of the ACS/WFC throughput curve in the near infrared 
(beyond 9000\AA) will be performed
during the calibration period of HST/Cycle 14 at STScI. The results of these test will most likely permit us to better quantify the uncertainties in the $z_{850}$ zero point. 


\subsection{Luminous mergers}

\begin{figure*}
\centerline{\includegraphics[angle=0,scale=0.7]{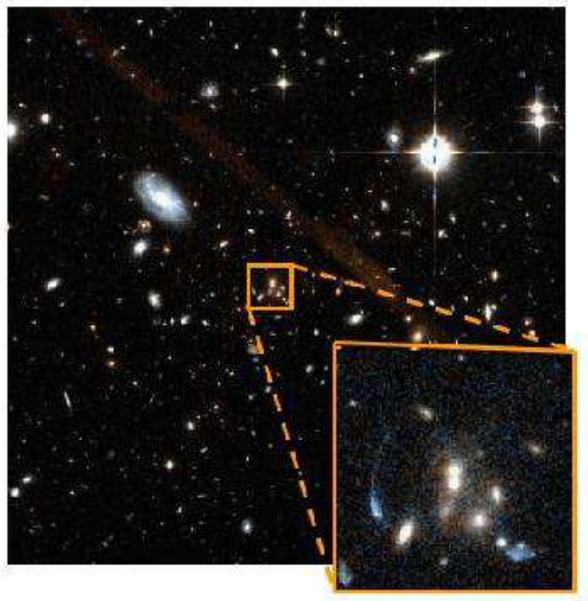}}
\caption {Lynx E ACS image (scale is 1~$\arcmin \times$ 1~$\arcmin$). The central on--going merger is magnified to also show
a gravitational arc and its likely counter image. {\label{cluster1}}}
\end{figure*}

Both Lynx E and Lynx W show evidence of on--going luminous mergers (van Dokkum 2001, Yamada et al. 2002). 
In the luminous merger close to the X--ray center of Lynx E (Fig.~\ref{cluster1}; zoomed image), three galaxies (the two ellipticals and the S0 to the north) were included in our color--magnitude relation, while the other two luminous galaxies were not, because they were classified as spirals. Their magnitudes range between $z_{850}=21.7$~mag and 22.7~mag, their $(i_{775} - z_{850})$ colors from $\approx$~0.97~mag to 1~mag. They all lie on the cluster CMR, as do the colors and magnitudes obtained by summing their fluxes. A single galaxy product of their merger would then also lie close to the CMR fit.  

\begin{figure*}
\centerline{\includegraphics[angle=-90,scale=0.7]{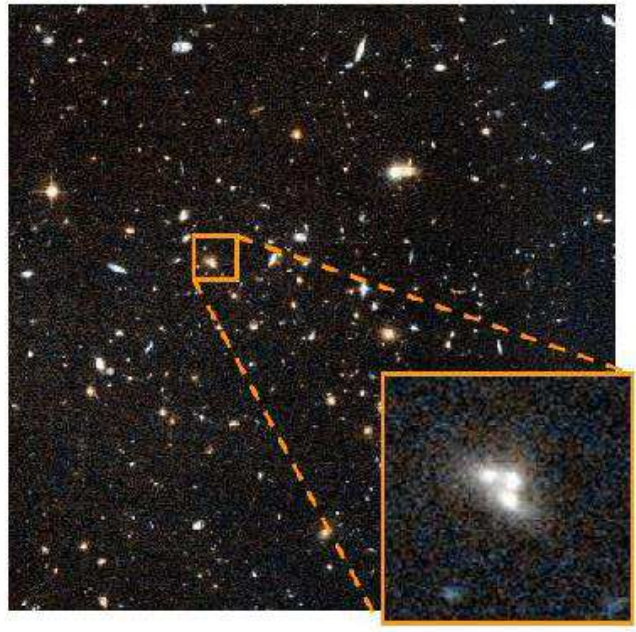}}
\caption {Lynx W ACS image (scale is 1~$\arcmin \times$ 1~$\arcmin$). The observed on--going merger is magnified. {\label{cluster2}}}
\end{figure*}

In our morphological classification, the galaxies in the luminous merger in Lynx W (Fig.~\ref{cluster2}; zoomed image) are
S0/a. The total Sextractor auto magnitude and colors of this last merger are shown as a green diamond in Fig~\ref{cmd}. It lies on the cluster red sequence.


\section{Discussion and conclusions}

The Lynx supercluster provides  a unique opportunity to study galaxy populations at an early stage of cluster formation. 
We have studied the color--magnitude relation and galaxy color distribution in the two clusters Lynx E and Lynx W.
Our results show that their elliptical populations are very similar, 
despite the fact that their different X--ray luminosities and shapes suggest that Lynx E is more evolved than Lynx W.

From the scatter around the cluster CMRs within 2$\arcmin$ of the cluster centers, and using a simple stellar population model from Bruzual and Charlot (2003),  mean luminosity--weighted ages of $\overline t > 2.5$~Gyr ($z_f > $~2.75) and $\overline t > 2.6$~Gyr ($z_f > $~2.8) were obtained for elliptical galaxies in Lynx E and Lynx W, respectively. Within 1$\arcmin$ from the cluster center, we found $\overline t > 3.4$~Gyr ($z_f > $~4.2) and $\overline t > 3.2$~Gyr ($z_f \approx$~3.7) for ellipticals in Lynx E and Lynx W, respectively.
CMR scatters decrease and CMR slopes steepen towards the cluster centers, suggesting the presence of younger ellipticals in the cluster outskirts, or more contamination by interlopers.
S0s and S0/as follow the same CMR as the ellipticals, and show younger mean ages than the ellipticals by $\sim$~0.5~Gyr, according to the scatter around the CMR.

\begin{figure*}
\centerline{\includegraphics[scale=0.7]{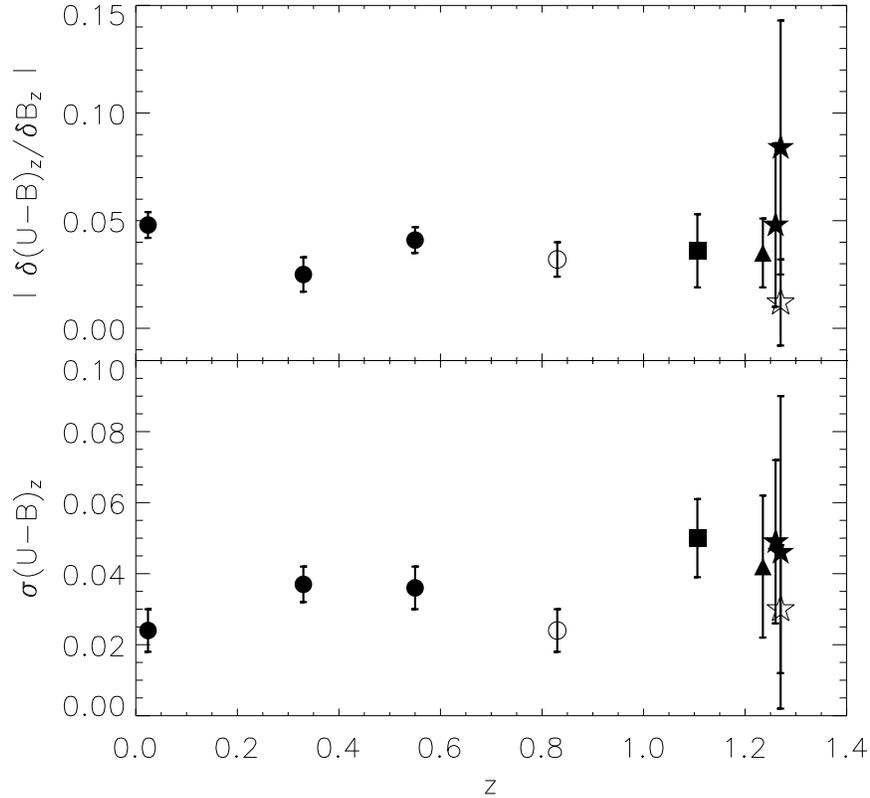}}
\caption {CMR absolute slope ($|\frac{\delta (U-B)_z}{\delta B_z}|$) and scatter ($\sigma  (U-B)_z$) for ellipticals as a function of redshift are shown by circles from Bower, Lucey, Ellis (1992) results for the Coma and Virgo clusters; van Dokkum et al. (1998) results for MS 1054-03; Ellis et al. (1997) results for a sample of nearby clusters of galaxies  (from left to right, in order of increasing redshift, as in Blakeslee et al. 2003b). A square shows Blakeslee et al. (2003) results for RDCS~1252--2927 and a triangle Mei et al. (2005) results for RXJ~J0910+5422. Stars shows combined results for Lynx E and Lynx W from this paper and  van Dokkum et al. (2001). van Dokkum et al. (2000) (MS 1054-03) and van Dokkum et al. (2001) (Lynx W) results concern all early--type galaxies and are shown by open symbols. These results do not indicate a significant dependence of absolute slopes on redshift, while scatters slightly increase with redshifts.  {\label{delta_ub}}}
\end{figure*}

Previous results from our Intermediate Redshift Cluster Survey, from Blakeslee et al (2003), showed that CMR scatters and slopes vary little with redshift.
Fig~\ref{delta_ub} shows elliptical CMR absolute slopes $|\frac{\delta (U-B)_z}{\delta B_z}|$ and scatters $\sigma (U-B)_z$ from the Bower, Lucey, Ellis (1992) results for the Coma and Virgo clusters; van Dokkum et al. (1998) results for CL 1358+62; Ellis et al. (1997) results for a sample of nearby clusters of galaxies; Blakeslee et al. (2003) results for RDCS~1252--2927; Mei et al. (2006) results for RXJ~J0910+5422; and this work  (Lynx E at z=1.26 and Lynx W at z=1.27).  van Dokkum et al. (2000) (MS 1054-03) and van Dokkum et al. (2001) (Lynx W) results for early--type galaxies are shown by open symbols. CMR slopes and scatter have been K--corrected to rest--frame $U-B$ color slopes and scatters using BC03 solar metallicity stellar population models. These studies probed the inner $\approx$~1~Mpc of the clusters.

A linear fit to the elliptical CMR slope and scatter gives $|\frac{\delta (U-B)_z}{\delta B_z}| = (0.043 \pm 0.005) - (0.009 \pm 0.010) z$, and $\sigma (U-B)_z = (0.027 \pm 0.005) + (0.020 \pm 0.009) z$, excluding from the fit the Lynx W elliptical CMR slope because of its large uncertainties. 
 While the slope does not show significant evolution as a function of redshift, consistent with the Blakeslee et al. (2003) results, the scatters within $\approx$~1~Mpc increase slightly with redshift. 
The Lynx W elliptical CMR slope  is steeper than the expected slope at z=1.27 (from the above fit $|\frac{\delta (U-B)}{\delta B}|_{z=1.27} = 0.052 \pm  0.010$), but still consistent within the large uncertainties.

As already pointed out by van Dokkum et al. (2001) and Yamada (2002) in their analyses of WFPC2/HST images of these clusters, two luminous red mergers are observed. The first one, in the very center of Lynx E, is composed of two ellipticals and one S0 that lie on the cluster CMR. The second one, in Lynx W, includes three S0/a galaxies, also lying on the cluster CMR. Qualitatively, the presence of these mergers confirms a scenario in which bright CMR galaxies are formed from hierarchical mergers of less luminous CMR galaxies.
Larger cluster samples at redshift larger than unity will permit us to establish if this is a general trend in high redshift clusters.

Other results from the ACS Intermediate Redshift Cluster Survey are supporting a view of galaxy cluster evolution in which, at redshift around unity, the cluster early--type population is still forming (Ford et al. 2004). While the old bright elliptical population (with a formation redshift around $z\approx3$; Blakeslee et al. 2003) is already evolved and has formed a tight CMR, there is an observed deficit of cluster S0s (Postman et al. 2005), that are probably not yet formed at z$\sim$~1. In one cluster, RXJ~J0910+5422 at z=1.11 (Mei et al. 2006), we observe a CMR S0 population that is bluer than the elliptical CMR, suggesting that we are witnessing a S0 population that is still evolving towards redder CMR elliptical colors. In the Lynx clusters, ellipticals and S0s all lie on the same tight CMR, with similar zero points and scatters for the two clusters, and similar slope for galaxies brighter than $M_B = -20.9 $~mag.
CMR scatters increase and CMR slopes are steeper with distance from the cluster center, suggesting younger populations in the cluster  cluster regions, or that there are more interlopers within 2~$\arcmin$. 
This similarity (at least for $M_B < -20.9$~mag) suggests that most of the bright older cluster elliptical population formed at the same time in both clusters, even if they look very different with respect to their X--ray emission shape and luminosity. Bright red sequence ellipticals were already in place before these two clusters evolve dynamically and merge into a larger structure.
Stanford et al. (1998) already pointed out that cluster galaxy colors do not depend on the cluster optical richness or X--ray luminosity.
Wake et al. (2005) found a similar result in a sample of 12 X--ray selected clusters spanning a large range in X--ray luminosities (and hence masses), from $L_X \approx  10^{43}$~erg~$s^{-1}$ to  $L_X \approx  10^{45}$~erg~$s^{-1}$ at z$\approx$~0.3.

Our results on these Lynx clusters are consistent with a scenario in which: 1) bright red ellipticals are already in place in clusters at redshift unity; 2) the color--magnitude relation is similar in clusters of different X--ray luminosities and dynamical state for galaxies with magnitudes $M_B$ brighter than -20.9~mag; 3) (at least part of the) bright ellipticals formed as a consequence of mergers of less luminous early--type but already red galaxies.

A more detailed study of galaxies as a function of environment will be possible with future planned ACS imaging of the groups surrounding the two main clusters studied in this paper. A comparison of bright and fainter galaxy properties in the clusters, groups and the field around them, will permit us to better understand the influence of the environment on their formation and evolution.

\begin{acknowledgements}
ACS was developed under NASA contract NAS 5-32865, and this research 
has been supported by NASA grant NAG5-7697 and 
by an equipment grant from  Sun Microsystems, Inc.  
The {Space Telescope Science
Institute} is operated by AURA Inc., under NASA contract NAS5-26555.
Some of the data presented herein were obtained at the W.M. Keck
Observatory, which is operated as a scientific partnership among the
California Institute of Technology, the University of California and
the National Aeronautics and Space Administration. The Observatory was
made possible by the generous financial support of the W.M. Keck
Foundation.  The authors wish to recognize and acknowledge the very
significant cultural role and reverence that the summit of Mauna Kea
has always had within the indigenous Hawaiian community.  We are most
fortunate to have the opportunity to conduct observations from this
mountain.
We are grateful to K.~Anderson, J.~McCann, S.~Busching, A.~Framarini, S.~Barkhouser,
and T.~Allen for their invaluable contributions to the ACS project at JHU. 
We thank W. J. McCann for the use of the FITSCUT routine for our color images.
\end{acknowledgements}

\newpage

\begin{table*}
\begin{center}
\caption{Lynx E cluster red sequence sample \label{rs_east}}
\vspace{0.25cm}
\resizebox{!}{7cm}{
\begin{tabular}{llccccccccccccccc}
\tableline \tableline\\
ACS ID & $z_{850}$ & $(i_{775} - z_{850})$&$R_e$&R&Morph \\
&(mag)&(mag)&$\arcsec$&(kpc)&\\
 \tableline\\
   E   0 &      21.87 $\pm$        0.04 &       1.00 $\pm$      0.01 &  0.214 &  13 &  E     \\
   E   1 &      22.00 $\pm$        0.04 &       1.03 $\pm$      0.01 &  0.271 &   8 &  E     \\
   E   2 &      22.04 $\pm$        0.02 &       0.99 $\pm$      0.01 &  0.529 & 123 & Sa     \\
   E   3 &      22.39 $\pm$        0.04 &       1.02 $\pm$      0.01 &  0.299 & 454 &  E     \\
   E   4 &      22.45 $\pm$        0.02 &       0.99 $\pm$      0.03 &  0.840 & 208 &  E     \\
   E   5 &      22.60 $\pm$        0.02 &       0.94 $\pm$      0.02 &  0.522 & 602 &  E     \\
   E   6 &      22.68 $\pm$        0.02 &       0.97 $\pm$      0.02 &  0.282 &  11 & S0     \\
   E   7 &      22.69 $\pm$        0.03 &       1.05 $\pm$      0.02 &  0.454 & 762 & S0     \\
   E   8 &      22.71 $\pm$        0.04 &       0.93 $\pm$      0.01 &  0.172 & 153 & S0     \\
   E   9 &      22.75 $\pm$        0.02 &       1.00 $\pm$      0.02 &  0.324 & 111 & S0     \\
   E  10 &      23.02 $\pm$        0.01 &       0.97 $\pm$      0.02 &  0.243 & 206 &  E     \\
   E  11 &      23.31 $\pm$        0.01 &       0.98 $\pm$      0.01 &  0.150 &  76 &  E     \\
   E  12 &      23.44 $\pm$        0.01 &       0.94 $\pm$      0.02 &  0.150 & 171 &  E     \\
   E  13 &      23.50 $\pm$        0.01 &       0.81 $\pm$      0.03 &  0.324 & 212 & Sa     \\
   E  14 &      23.50 $\pm$        0.02 &       1.06 $\pm$      0.02 &  0.150 & 693 & S0     \\
   E  15 &      23.58 $\pm$        0.02 &       0.98 $\pm$      0.02 &  0.244 &  71 &  E     \\
   E  16 &      23.60 $\pm$        0.01 &       0.92 $\pm$      0.03 &  0.261 & 237 & S0     \\
   E  17 &      23.63 $\pm$        0.02 &       0.89 $\pm$      0.12 &  0.953 & 561 & Sa     \\
   E  18 &      23.66 $\pm$        0.01 &       1.05 $\pm$      0.02 &  0.150 & 705 &  E     \\
   E  19 &      23.69 $\pm$        0.03 &       0.96 $\pm$      0.02 &  0.207 &1000 & S0     \\
   E  20 &      23.73 $\pm$        0.01 &       1.00 $\pm$      0.04 &  0.365 & 242 & S0     \\
   E  21 &      23.77 $\pm$        0.03 &       0.90 $\pm$      0.03 &  0.294 & 152 & S0     \\
   E  22 &      23.82 $\pm$        0.02 &       1.02 $\pm$      0.39 &  4.259 & 301 & Sa     \\

\tableline \tableline\\

\end{tabular}}
\end{center}

\end{table*}

\begin{table*}
\begin{center}
\caption{Lynx W cluster red sequence sample \label{rs_west}}
\vspace{0.25cm}
\resizebox{!}{6cm}{
\begin{tabular}{llcccccccccccc}
\tableline \tableline\\
ACS ID & $z_{850}$ & $(i_{775} - z_{850})$&$R_e$&R&Morph \\
&(mag)&(mag)&$\arcsec$&(kpc)&\\
\tableline \\
   W   0 &      21.72 $\pm$        0.04 &       1.02 $\pm$      0.10 &  2.347 & 135 &  E     \\
   W   1 &      21.99 $\pm$        0.04 &       0.99 $\pm$      0.02 &  0.915 & 179 & Sa     \\
   W   2 &      22.14 $\pm$        0.02 &       0.99 $\pm$      0.02 &  0.538 & 373 &  E     \\
   W   3 &      22.46 $\pm$        0.04 &       1.06 $\pm$      0.01 &  0.236 & 156 & S0     \\
   W   4 &      22.49 $\pm$        0.02 &       1.00 $\pm$      0.01 &  0.693 & 178 & Sa     \\
   W   5 &      22.53 $\pm$        0.02 &       0.99 $\pm$      0.01 &  0.280 & 886 & Sa     \\
   W   6 &      22.72 $\pm$        0.02 &       1.00 $\pm$      0.02 &  0.453 & 158 & Sa     \\
   W   7 &      23.04 $\pm$        0.03 &       0.99 $\pm$      0.02 &  0.390 & 233 &  E     \\
   W   8 &      23.11 $\pm$        0.04 &       0.98 $\pm$      0.05 &  0.719 & 224 & Sa     \\
   W   9 &      23.15 $\pm$        0.02 &       1.00 $\pm$      0.02 &  0.253 & 807 &  E     \\
   W  10 &      23.16 $\pm$        0.01 &       0.90 $\pm$      0.04 &  0.689 &  48 &  E     \\
   W  11 &      23.26 $\pm$        0.01 &       1.08 $\pm$      0.01 &  0.167 & 664 & Sa     \\
   W  12 &      23.36 $\pm$        0.01 &       0.96 $\pm$      0.02 &  0.256 & 870 & Sa     \\
   W  13 &      23.56 $\pm$        0.01 &       0.95 $\pm$      0.02 &  0.189 & 296 & S0     \\
   W  14 &      23.81 $\pm$        0.02 &       0.94 $\pm$      0.02 &  0.150 & 128 &  E     \\
   W  15 &      23.83 $\pm$        0.02 &       0.90 $\pm$      0.04 &  0.360 & 574 & S0     \\
   W  16 &      23.84 $\pm$        0.01 &       0.85 $\pm$      0.05 &  0.329 & 918 & Sa     \\

\tableline \tableline\\

\end{tabular}}
\end{center}

\end{table*}

\begin{table*}
\begin{center}
\caption{Color--Magnitude Relations \label{results}}
\vspace{0.25cm}
\resizebox{!}{10cm}{
\begin{tabular}{llcccccccccccc}
\tableline \tableline\\
Cluster & Sample &$N$&c$_0$&$Slope$ & $\sigma_{int}$\\
&&&(mag)&&(mag)&&\\
\tableline \tableline\\
 Lynx E$^1$ &E+S0+S0/a & 17 &       0.99  $\pm$       0.01 &     -0.038  $\pm$    0.020 &    0.040  $\pm$    0.025     \\
                         &E+S0 & 14 &       0.99  $\pm$       0.01 &     -0.040  $\pm$    0.015 &    0.020  $\pm$    0.006     \\
                            &E &  8 &       1.00  $\pm$       0.01 &     -0.032  $\pm$    0.012 &    0.011  $\pm$    0.007     \\
    Lynx W$^1$ &E+S0+S0/a & 11 &       0.99  $\pm$       0.01 &     -0.038  $\pm$    0.019 &    0.023  $\pm$    0.013     \\
                         &E+S0 &  7 &       0.99  $\pm$       0.02 &     -0.052  $\pm$    0.021 &    0.031  $\pm$    0.019     \\
                            &E &  5 &       0.97  $\pm$       0.01 &     -0.053  $\pm$    0.023 &    0.017  $\pm$    0.028     \\
            Both$^1$&E+S0+S0/a & 28 &       0.99  $\pm$       0.01 &     -0.039  $\pm$    0.013 &    0.034  $\pm$    0.015     \\
                         &E+S0 & 21 &       0.99  $\pm$       0.01 &     -0.046  $\pm$    0.010 &    0.023  $\pm$    0.007     \\
                     &E+S0$^a$ & 13 &       1.00  $\pm$       0.01 &     -0.031  $\pm$    0.025 &    0.019  $\pm$    0.011     \\
                            &E & 13 &       0.99  $\pm$       0.01 &     -0.036  $\pm$    0.011 &    0.017  $\pm$    0.010     \\
                           &S0 &  8 &       1.00  $\pm$       0.03 &     -0.063  $\pm$    0.035 &    0.032  $\pm$    0.013     \\

\tableline\\

 Lynx E$^2$ &E+S0+S0/a & 21 &       0.98  $\pm$       0.01 &     -0.026  $\pm$    0.020 &    0.050  $\pm$    0.021     \\
                         &E+S0 & 17 &       0.99  $\pm$       0.01 &     -0.014  $\pm$    0.021 &    0.038  $\pm$    0.008     \\
                            &E & 10 &       0.99  $\pm$       0.01 &     -0.025  $\pm$    0.020 &    0.026  $\pm$    0.012     \\
    Lynx W$^2$ &E+S0+S0/a & 13 &       1.00  $\pm$       0.01 &     -0.045  $\pm$    0.016 &    0.030  $\pm$    0.014     \\
                         &E+S0 &  8 &       0.99  $\pm$       0.02 &     -0.053  $\pm$    0.018 &    0.027  $\pm$    0.017     \\
                            &E &  5 &       0.97  $\pm$       0.01 &     -0.053  $\pm$    0.023 &    0.017  $\pm$    0.028     \\
            Both$^2$&E+S0+S0/a & 34 &       0.99  $\pm$       0.01 &     -0.035  $\pm$    0.013 &    0.044  $\pm$    0.013     \\
                         &E+S0 & 25 &       0.99  $\pm$       0.01 &     -0.032  $\pm$    0.013 &    0.037  $\pm$    0.008     \\
                     &E+S0$^a$ & 14 &       1.00  $\pm$       0.01 &     -0.021  $\pm$    0.030 &    0.023  $\pm$    0.012     \\
                            &E & 15 &       0.99  $\pm$       0.01 &     -0.032  $\pm$    0.014 &    0.025  $\pm$    0.010     \\
                           &S0 & 10 &       1.00  $\pm$       0.03 &     -0.051  $\pm$    0.038 &    0.046  $\pm$    0.015     \\

\tableline\\
    Lynx E$^3$ &E+S0+S0/a & 23 &       0.99  $\pm$       0.01 &     -0.026  $\pm$    0.017 &    0.049  $\pm$    0.019     \\
                         &E+S0 & 19 &       0.99  $\pm$       0.01 &     -0.020  $\pm$    0.018 &    0.038  $\pm$    0.008     \\
                            &E & 10 &       0.99  $\pm$       0.01 &     -0.025  $\pm$    0.020 &    0.026  $\pm$    0.012     \\
      Lynx W$^3$ &E+S0+S0/a & 17 &       1.00  $\pm$       0.01 &     -0.051  $\pm$    0.019 &    0.039  $\pm$    0.014     \\
                         &E+S0 &  9 &       1.00  $\pm$       0.02 &     -0.056  $\pm$    0.018 &    0.027  $\pm$    0.015     \\
                            &E &  6 &       0.98  $\pm$       0.02 &     -0.043  $\pm$    0.031 &    0.024  $\pm$    0.023     \\
              Both$^3$&E+S0+S0/a & 40 &       0.99  $\pm$       0.01 &    -0.039  $\pm$    0.012 &    0.046  $\pm$    0.011     \\
                         &E+S0 & 28 &       1.00  $\pm$       0.01 &     -0.035  $\pm$    0.012 &    0.037  $\pm$    0.007     \\
                     &E+S0$^a$ & 10 &       1.01  $\pm$       0.02 &     -0.005  $\pm$    0.044 &    0.033  $\pm$    0.023     \\
                            &E & 16 &       0.99  $\pm$       0.01 &     -0.031  $\pm$    0.012 &    0.025  $\pm$    0.010     \\
                           &S0 & 12 &       1.02  $\pm$       0.03 &     -0.056  $\pm$    0.031 &    0.043  $\pm$    0.011     \\

\tableline \tableline
\end{tabular}}
\end{center}

\end{table*}

\clearpage

\end{document}